\documentclass[12pt]{article}

\usepackage{epsf}

\usepackage{graphicx}
\begin{document}

Astrophysics, Vol. 53, No. 4, 2010

\begin{center}

{\bf \Large ISOLATED DWARF GALAXIES IN THE LOCAL SUPERCLUSTER AND ITS SURROUNDINGS}\\

\bigskip

{\large V. E. Karachentseva$^1$, I. D. Karachentsev$^2$, and M. E. Sharina$^2$} \\

\bigskip

($^1$) Main Astronomical Observatory, National Academy of Sciences of Ukraine, Ukraine; e-mail: valkarach@gmail.com \\
($^2$) Special Astrophysical Observatory, Russian Academy of Sciences; e-mail: ikar@sao.ru\\

\bigskip

{\bf Abstract} \\ 

\end{center}

We present a list of 75 isolated late-type dwarf galaxies which have no neighbors with a relative radial velocity difference of less than 500 km/s or projected separations within  500 kpc. These were selected from $\sim$2000 dwarf galaxies with radial velocities $V_{LG}<3500$ km/s within the volume of the Local supercluster. In terms of their sizes, luminosities, and the amplitudes of their internal motions, the isolated late-type dwarfs  do not differ significantly from gas-rich dwarf galaxies in groups and clusters. However, the median mass of neutral hydrogen per unit luminosity for the isolated dwarf galaxies is two times more  than that for the late-type galaxies in groups. We have also identified 10 presumably isolated spheroidal dwarf galaxies. The detection  of isolated dwarf galaxies populated exclusively by old stars is of great interest for modern cosmological scenarios of galaxy formation.

\section{Introduction}
Isolated galaxies are studied more and more actively since recent years. (See, for example, the Proceedings of the Conference : ``Galaxies in Isolation: Exploring nature vs. Nurture'' [1]). Studies of galaxies in low density regions  allow to investigate the influence of environments on star formation processes and the evolution of galaxies, and to test various models of the origin of galaxies. Isolated dwarf galaxies, including those with low surface brightness, are of special interest for these problems.

The typical characteristics of dwarf galaxies (linear sizes of a few kpc, absolute $B$ magnitudes fainter than $-16.5^m$, mean surface brightness $<SB>$ fainter than 25 B mag/arcsec$^2$) [2] explain the difficulty of detecting them at large distances. Thus, dwarf galaxies are poorly represented in catalogs limited by apparent  magnitude (CGCG [3], $\leq 15.7^m$ ) or by angular diameter,  $a\geq 1^{\prime}$ (UGC [4], ESO [5]). Most objects in the UGC catalog that have been classified as ``dwarfs'' by their low surface brightness turned out to be ordinary spiral galaxies after their radial velocities were measured. The relative number of true dwarf galaxies in the UGC [4] is only 2\%. There are also few dwarf galaxies in the catalog of isolated galaxies compiled by Karachentseva [6] on the basis of the CGCG catalog. Detection of low surface brightness dwarf galaxies is an extremely time-consuming task. 

The situation regarding the representation of dwarf systems changes drastically  when the sample of galaxies is limited by distance, rather than flux. Thus, the Catalog of Neighboring Galaxies [2] includes 451 objects in the Local volume with distances of up to 10 Mpc; of these about 85\% of the galaxies are dwarf systems, mostly with low and very low surface brightness. A determination of the so-called ``tidal index'' [2] for each galaxy ``i'' in the Local volume,

       $$\Theta=\max[\log(M_k/D^3_{ik})], \,\, k=1,2,\ldots$$
where $M_k$ is the mass and $D_{ik}$ is the mutual spatial distances of neighboring galaxies, has shown that about half the population of the Local volume occurs in groups of varying multiplicities. The rest of them are gravitationally unbound with a negative tidal index ($\Theta  < 0)$ and can be treated as objects in the general field. The most isolated of them, with  $\Theta < -2$ , turned out to be dwarf systems almost exclusively (16 out of 18).

 A similar approach for identifying isolated galaxies has been developed in [7]. A bounding algorithm was applied to a sample of 10914 galaxies with radial velocities in the Local Group frame, $V_{LG} < 3500$ km/s. After pairs, triplets, and groups had been identified [8-10], roughly 46\% of the remaining galaxies appeared to be unclusterized or, as they are often called, field galaxies. A more rigorous criterion of isolation, in particular the criterion [6], left only about 500 very isolated galaxies in this volume [11]. Roughly 13\% of them are objects with low surface brightnesses. 

An average relative number of dwarf galaxies with respect to normal ones is larger in the Local volume ($V_{LG} < 550$ km/s), than in the Local supercluster ($V_{LG} < 3500$ km/s). This observational selection effect
  is related both to the obvious decrease in the number of detected faintest dwarf systems with increasing distance, and to the fact that less attention has been paid to searches for dwarfs in the field than in the well-known nearby groups and clusters of galaxies. Nevertheless, the overall distribution of dwarf galaxies in the Local supercluster and its surroundings has been determined quite reliably up to the date. They are concentrated mainly in the Virgo and Fornax clusters and in systems having different populations, with spheroidal dwarfs located in the dense central parts and late-type dwarfs , in the more rarefied peripheral regions of groups and clusters. 

This paper is the result of a search for fairly isolated dwarf galaxies of different morphological types within the volume $V_{LG} < 3500$ km/s. We have examined a combined sample of known field dwarf galaxies and have attempted to identify the most isolated objects within that sample. While not claiming statistical completeness of the sample, we hope to draw the attention of observers to these objects and to compare the characteristics of dwarf galaxies whose
isolation has been established by various criteria.

\section{Initial data and sampling criteria}

From the complete set of known dwarf galaxies within the volume $V_{LG} < 3500$ km/s, we have complied an overall summary of candidate galaxies which do not enter visually in  the Virgo and Fornax clusters, or known groups, and are not satellites of normal galaxies. The following sources were used for this purpose:

 (1) A catalog of 1500 low surface brightness dwarf galaxies (LSB) $(<SB>$ fainter than 24.5 $B$ mag/ arcsec$^2$) [12]. This catalog was the first attempt to gather, test, and generalize the results of searches for and observations of dwarf galaxies in articles published by various authors up to 1987. This catalog covers the entire sky; its average depth is $\sim$1500 km/s and roughly 500 objects appear in the common field. 

(2) New lists of LSB dwarf galaxies and articles about observations of these objects published after 1987, in particular searches for dwarf galaxies in the general field [13--16],  HI-observations of the UGC  dwarfs [17,18], and several others. Out of this set of data we verified roughly 250 objects.

 (3) Articles presenting the results of continuous systematic searches for dwarf galaxies in the photographic sky surveys POSS-II (KK, KKR, KKH) [19-21], ESO-SERC (KKs) [22] and ESO SERC-J (KKSG)[23]  the ``5 lists'', with a total number $N\sim$600. Observations of these objects in the HI 21 cm line [21--27] indicated a fairly high level of detection, up to 60\%. Roughly 300 objects were verified from these sources. 

 We excluded from the lists of the candidates the objects of our Galaxy: planetary nebulae, reflecting nebulae, cirri, and also dwarf galaxies with high surface brightness. Galaxies satisfying the following conditions were chosen for further testing:

 (1) objects which have radial velocities with respect to the center of the Local group within the range $500 < V_{LG} <3500$ km/s and do not belong to groups and clusters; 

 (2) dwarf galaxies found outside the zone of strong Galactic absorption, i.e., has $A_B < 1^m$.
 
We did not examine dwarf galaxies within the Local volume, since their properties have been summarized in the CNG catalog [2];

After refinement of the coordinates, verification of mutual identifications in the lists, and elimination of common objects independently included in different sources, we obtained a refined list of 274 late-type dwarf galaxies  with measured radial velocities. Each of them was then tested according to the criterion of isolation. Here, we consider a dwarf galaxy "1" to be  isolated if it has no neighbors ``i'' with relative radial velocities $\mid V_{1i}\mid < 500$ km/s and projected distances $R_p < 500$ kpc.

 We use data on the coordinates and radial velocities of the surrounding galaxies from the NED database, using it to determine the distances to nearest neighbors. No limits were imposed on the  magnitude of the neighboring galaxies. The sample of very isolated galaxies in the Local volume and their main characteristics are discussed in section 5. We checked the dwarf spheroidal galaxies without radial velocity estimates separately (section 4). Seventy five of the 274 late-type dwarf galaxies (i.e. 27\%) satisfied the isolation criterion. Of these 15 have no nearest neighbors out  to a distance of 1 Mpc.

\section{List and characteristics of isolated late-type dwarf galaxies} 

Table 1 lists some of the observational data for the late type isolated dwarf galaxies. These data are mainly taken from the HyperLEDA database, and when they are lacking, from the NED database, or from original articles. In calculating the main optical and HI characteristics, we have used the computational scheme of  [2] with the Hubble constant taken to be $H_0$ = 73 km/s/Mpc. Columns 1--10 of Table 1 list the following: (1) the galaxy name; (2) the equatorial coordinates at epoch J2000.0; (3) the morphological type determined by us from images of the galaxies on the sky surveys POSS II, ESO/ SERC, DSS-1, red, blue, and, where possible, SDSS (in column 3 we omit the letter ``d'', for dwarf). In this paper we have divided the low surface brightness dwarf galaxies in accordance with the classification of  [12]: Sm --- galaxies with a disrupted spiral structure, but with a distinct center and, in some cases, with residual signs of a bar; Im --- diffuse galaxies with ``magellanic'' features, randomly positioned blue knots ; Ir --- diffuse galaxies, also of an irregular shape, but without visible blue knots ; and, Sph --- diffuse galaxies with a regular circular or elliptical shape. (4) The apparent B magnitude taken from HyperLEDA and, in some cases, from NED or original papers. Note that the error in the estimated stellar magnitude for these dwarfs can be on the order of 0.5$^m$ or worse. (5) Radial velocity in km/s reduced to the centroid of the Local group with the apex parameters accepted in NED. (6) The maximum value of the rotation velocity in km/s, corrected for the inclination, taken from HyperLEDA. (7) The absolute B magnitude corrected for Galactic and internal absorption. (8) The logarithm of the hydrogen mass, in solar units. (9) The logarithm of the hydrogen mass -to- $B$-luminosity ratio,in solar units. (10) The most isolated dwarf galaxies with no neighboring galaxies at distances to $R_p$ = 1 Mpc are indicated by an asterisk (*). 

The last row of the table lists the median values of these characteristics of the galaxies. The following notes for Table 1 describe the shapes of some of these objects on the SDSS survey images.

\begin{table}        
\caption{List of isolated late-type dwarf galaxies  within the Volume $V_{LG} < 3500$ km/s}
\begin{tabular}{lcllrrrrrc}\\ \hline

  Name          & RA(J2000.0)DEC  & Type    & $B$&    $V_{LG}$& $V_r$& $M_B^c$&  $\log(M_{HI})$& $\log(M_{HI}/L_B)$& * \\
   \hline

  (1)           &    (2)       &  (3) & (4) & (5) &(6) &(7) & (8)  &     (9) &  (10)\\
\hline
 ESO 12-14       &000244.2-802048& Sm   &  14.83 & 1757& 44 &-17.5&  9.81 &   0.65&   \\
 UGC 199         &002051.8+125122& Sm   &  17.3  & 2012& 46 &-15.2&  8.78 &   0.54&   \\
 ESO 474-25      &004707.5-223542& Sm   &  16.0  & 2896& 57 &-17.1&  9.12 &   0.14&  *\\
 UGC 672         &010617.9+445715& Ir   &  17.1  &  967& 36 &-14.1&  7.89 &   0.11&   \\
 ESO243-50       &011048.8-422231& Im/Ir&  17.04 & 1413& 26 &-14.4&  8.78 &   0.86&  *\\
 LSBC F683-V10   &013031.9+112925& Ir   &  17.51 & 1134& -  &-13.9&   -   &    -  &   \\
 KDG 9           &014234.0+025546& Ir   &  16.9  & 1882& -  &-15.3&   -   &    -  &   \\
 KKH 8           &021227.4+101958& Ir   &  16.7  & 1861& 24 &-15.8&  8.30 &  -0.20&   \\
 ESO 545-2       &021915.3-185556& Sm   &  14.96 & 1603& 58 &-17.1&  9.14 &   0.15&   \\
 ESO 298-33      &022128.1-384803& Im   &  17.76 & 2142& 25 &-14.7&  8.81 &   0.79&   \\
 ESO 545-15      &022557.9-194130& Im   &  16.72 & 2272& 33 &-15.9&  9.37 &   0.86&   \\
 KKH 14          &024502.9+320942& Ir   &  17.0  & 1748& 45 &-15.7&  9.27 &   0.82&  *\\
 UGC 2352        &025205.4+042215& Im   &  17.2  & 1866& 35 &-15.1&  8.86 &   0.67&   \\
 KDG 32          &031856.7-103247& S0   &  16.74 & 1898& 42 &-15.7&  9.04 &   0.60&  *\\
 AM 0333-611     &033415.3-610548& Im   &  16.38 &  976& 24 &-14.4&  7.96 &   0.04&   \\
 HIPASS J0341+18 &034201.7+180831& Im   &  16.90 & 1369& 18 &-15.4&  8.37 &   0.06&   \\
 ESO 251-3       &042841.2-461916& Im   &  16.19 & 1191& 31 &-14.9&  8.32 &   0.19&   \\
 KK 271          &045145.3+670932& Ir/Sph& 17.7  & 1771& -  &-15.1&  8.04 &  -0.17&  *\\
 ESO 555-39      &060901.1-214323& Sm   &  15.89 & 1535& 59 &-16.1&  8.84 &   0.24&   \\
 UKS 0616-708    &061549.2-705342& Sm   &  15.5  & 1025& 19 &-15.6&  8.86 &   0.45&   \\
 KK 56           &064311.6+363803& Ir   &  17.9  & 2360& 29 &-15.3&  8.30 &   0.02&   \\
 UGC 3672        &070627.5+301919& Im/Pec& 16.4  &  969& 44 &-14.5&  9.10 &   1.12&   \\
 KK 58           &070910.7-512801& Ir   &  15.38 &  772& 17 &-15.3&  8.27 &  -0.01&   \\
 KK 66           &074729.8+401116& Ir   &  17.5  & 2949& 31 &-15.7&  8.85 &   0.39&   \\
 UGC 4117        &075726.0+355621& Im   &  15.85 &  754& 28 &-14.4&  8.08 &   0.15&   \\
 UGC 4100        &080656.9+844515& Im   &  16.0  & 1221& 38 &-15.4&  8.59 &   0.29& * \\
 KKH 44          &081638.5+692049& Ir   &  18.2  & 1202&  8 &-13.0&  7.88 &   0.51& * \\
 UGC 4527        &084424.0+765505& Ir   &  16.9  &  900& 68 &-13.8&  7.96 &   0.30&   \\
 KDG 54          &092225.2+754557& Im/Ir&  16.4  &  833& 20 &-14.0&  7.97 &   0.21&   \\
 KKH 52          &093747.2+273404& Im/Ir&  16.6  & 1513& 15 &-15.1&  8.45 &   0.27&   \\
 KDG 69          &102852.6+664823& Ir   &  17.4  & 1268& 33 &-13.8&  8.45 &   0.75&   \\
 UGC 5996        &105257.4+402242& Sm   &  16.8  & 1617& 37 &-15.1&  8.56 &   0.35&   \\
 UGC 6222        &111104.5+343411& Sm   &  16.2  & 1926& 33 &-16.0&  8.82 &   0.26&   \\
 KK 102          &112258.9+192839& Ir   &  16.45 & 3268& 45 &-16.9&  8.87 &  -0.05&   \\
 KK 105          &112924.2+460651& Ir   &  16.6  & 1598& 40 &-15.2&  8.36 &   0.13&   \\
 UGC 6996        &120022.7+785103& Sm   &  17.1  & 2161& 56 &-15.6&  8.75 &   0.36&   \\
 KKs  48          &120535.9-434612& Ir   &  17.62 & 2732& 45 &-15.8&  9.39 &   0.91&   \\
 KDG 89          &121428.9-121957& Ir   &  17.25 & 1714& 43 &-14.8&  8.57 &   0.49&   \\
 UGC 7995        &124957.7+782302& Ir   &  16.25 & 1998& 57 &-16.2&  8.59 &  -0.04& * \\
KK 181          &130433.7+264627& Im/Ir&  17.25 & 1914& 36 &-14.9&  8.77 &   0.65& * \\
 KK 183          &130642.4+180007& Sph/Ir& 17.9  & 1496& 43 &-13.8&  8.34 &   0.67&   \\
 UGC 8474        &132926.4+005412& Sm   &  16.26 & 3164& 30 &-17.0&  8.86 &  -0.11&   \\
 KDG 227         &133439.7-121950& Sm   &  15.59 & 1347& 41 &-16.0&  8.23 &  -0.31&   \\
 LSBC D721-V10   &135557.8+085950& Ir   &  17.8  & 1165& 22 &-13.3&  7.17 &  -0.32&   \\
 KK 226          &135608.4-453934& Sm   &  15.91 & 2300& 28 &-16.9&  9.06 &  -0.11&   \\
 KKR  2           &140626.2+092133& Ir/Im&  17.32 & 3213& 34 &-16.0&  8.76 &   0.19& * \\\end{tabular}
\end{table}

\begin{table}
\begin{tabular}{lcllrrrrrc}\\ \hline

   \hline

  (1)           &    (2)       &  (3) & (4) & (5) &(6) &(7) & (8)  &     (9) &  (10)\\
\hline
 
 UGC 9123        &141503.7+362726& Sm   &  15.73 & 1951& 62 &-16.5&  8.77 &   0.01&   \\
 KKR  6           &141703.6-013015& Im   &  16.93 & 1463& 25 &-14.8&  8.41 &   0.33&   \\
 KKSG 47         &143525.4-171001& Ir   &  17.7  & 1447& 54 &-14.2&  8.92 &   1.08&   \\
 KKR  12          &144624.3+141245& Ir   &  16.6  & 1801& 20 &-15.4&  8.77 &   0.43&   \\
 UGC 9875        &153047.3+230358& Sm   &  15.83 & 2073& 86 &-16.7&  8.95 &   0.12&   \\
 UGC 9912        &153510.5+163258& Sm   &  15.56 & 1032& 19 &-15.4&  8.60 &   0.28&   \\
 KKR  21          &153700.6+204742& Sm/Im&  16.8  & 1798& 34 &-15.4&  8.94 &   0.62&   \\
 UGC 10009       &154537.5+041057& Ir   &  17.0  & 2095& 64 &-16.0&  8.71 &   0.15&   \\
 UGC 10229       &160943.9-000654& Im   &  17.35 & 1522& 44 &-14.8&  9.15 &   1.08&   \\
 KKR  26          &161644.6+160509& Ir   &  17.1  & 2347& 24 &-15.6&  8.75 &   0.34& * \\
 UGC 10376       &162250.9+652616& Sm   &  16.5  & 3246& 36 &-16.8&  9.01 &   0.11&   \\
 LSBC F585-V01   &162557.4+203934& Ir   &  18.4  & 2106& 26 &-14.2&   -   &    -  &   \\
 KKR  30          &165638.5+075956& Ir   &  17.0  & 1584& 31 &-15.2&  8.73 &   0.49& * \\
 KKR  34          &171242.1+135428& Ir   &  17.8  & 1640& 37 &-14.6&  8.69 &   0.68&   \\
 KKR  39          &175900.7+215053& Ir   &  18.1  & 2242& 24 &-14.8&  8.76 &   0.67&   \\
 KKR  42          &181052.1+371453& Ir   &  17.7  & 1754& 12 &-14.4&  8.06 &   0.14&   \\
 KKs  68          &182246.5-621613& Im   &  16.7  &  808& 23 &-13.9&  7.81 &   0.10&   \\
 UGC 11458       &193506.9+695942& Ir   &  17.5  & 1651& 65 &-15.1&  8.54 &   0.36& * \\
 KKR  48          &195756.4+623721& Ir   &  17.5  & 3453& 17 &-16.2&  8.94 &   0.29&   \\
 KK 249          &202906.9-314110& Ir   &  15.79 & 2169& 66 &-17.1&  8.80 &  -0.21&   \\
 ESO 403-36      &215010.9-354228& Ir   &  16.90 & 2570& 75 &-16.1&  9.58 &   0.99&   \\
 KK 255          &215755.8-601822& Im   &  14.6  & 1585& 22 &-17.2&  8.11 &  -0.95&   \\
 ESO 189-21      &220236.6-540443& Sm   &  15.31 & 1644& 28 &-16.5&  8.94 &   0.16&   \\
 ESO 602-16      &222323.9-180730& Im   &  15.68 & 2723& 52 &-17.4&  9.26 &   0.16&   \\
 KKR  71          &223038.0+384355& Ir   &  17.3  & 1002& 11 &-14.0&  8.65 &   0.91&   \\
 UGC 12212       &225030.3+290818& Sm   &  16.2  & 1177& 45 &-15.1&  8.88 &   0.66&   \\
 KKR  75          &232011.2+103723& Ir   &  18.0  & 1703& 45 &-14.1&  8.66 &   0.87& * \\
 UGC 12070       &233157.3+780903& Im   &  16.5  & 1724& 48 &-16.3&  9.08 &   0.38& * \\
 UGC 12771       &234532.7+171512& Im/Ir&  16.5  & 1535& 38 &-15.4&  8.48 &   0.17&   \\
\hline
 MEDIANS     &&          &                 16.8  & 1720 &35 &-15.3 &8.76 &  0.30\\
\hline

\multicolumn{10}{l}{Notes for Table 1:} \\
\multicolumn{10}{l}{ESO 545-2: wedge shaped, with blue knots;}\\
 \multicolumn{10}{l}{UGC 3672: long diffuse ``tail'' with knots directed to the NW; }\\
\multicolumn{10}{l}{UGC 5996: distinct brightness gradient, a diffuse blue shell and several jets with knots;}\\
 \multicolumn{10}{l}{UGC 6222: possibly dIr, bluish, with diffuse knots;}\\
 \multicolumn{10}{l}{KK 181: blue, with diffuse ``spots'', without bright knots ; }\\
\multicolumn{10}{l}{UGC 8474: smooth spiral with a diffuse central part;}\\
 \multicolumn{10}{l}{LSBC D721-V10: noticeable brightness gradient, a star is projected;}\\
 \multicolumn{10}{l}{KKR  2: faint knots; }\\
\multicolumn{10}{l}{KKR  12: diffuse patchy knots;}\\
 \multicolumn{10}{l}{KKR  6: faint knots; }\\
\multicolumn{10}{l}{UGC 9123: regular diffuse spiral arms with several blue knots; }\\
\multicolumn{10}{l}{UGC 9875: regular diffuse spiral arms  with several blue knots;}\\
 \multicolumn{10}{l}{UGC 9912: regular spiral, diffuse nuclear part and blue knots in the spiral arms; }\\
\multicolumn{10}{l}{KKR  21: disrupted
spiral, very faint knots, a star is projected; }\\
\multicolumn{10}{l}{UGC 10009: irregular, without a brightness gradient;}\\
 \multicolumn{10}{l}{UGC 10229:  knots;}\\
 \multicolumn{10}{l}{KKR  26: faint blue knot;}\\
\multicolumn{10}{l}{ UGC 12771: diffuse, filamentary.}\\
\end{tabular}
\end{table}
\section{Candidate isolated dwarf spheroidal galaxies}

A search for presumed isolated spheroidal dwarfs in the field of the Local supercluster is a much more difficult task than searching for dwarfs of late types. By definition, spheroidal dwarfs do not contain a significant fraction of gas, so that they are not detected in the 21 cm line. Their optical velocities are hard to determine because of the low surface brightness of dSphs. In the general field beyond  the Local volume, as well as outside the Virgo and Fornax clusters, their distances were not determined directly. Spheroidal dwarfs are often morphologically similar to irregular objects, but they have more regular shape and a smoother structure. We tried to identify candidate spheroidal dwarf objects in the general field by their morphology and then, using indirect features, to distinguish which of them are isolated. 

We selected objects which have not been detected in the HI 21 cm line in the catalogs and lists cited in section 2, and classified  as dSph and/or dSph/dIr. Our new classification using the sky surveys DSS-1 and DSS-2, and, where possible SDSS, showed that the previous search, especially in the POSS-1 survey, often gave an erroneous type estimation, and this is further confirmed by the 21 cm line observations. Out of the several hundred objects, we selected 107 candidates for spheroidal dwarfs or dSph/dIr without HI radial velocities. Then, using the NED database for each object, we searched for neighbors within a wide range of radial velocities, 500--3500 km/s. It turns out that most of the candidates have neighboring galaxies with radial velocities on the order of 1000--2000 km/s. Taking the same values of the radial velocities for the candidate galaxies, as well, we obtained linear dimensions and absolute magnitudes of the dwarfs which are typical of dSph, and projected distance of $\sim$100--300 kpc for the neighbors. 
Under the assumed conditions, these dwarfs cannot be regarded as isolated, although they are, most likely, within the volume of the Local supercluster.

After verification of the 107 candidates in our list of presumed isolated dwarf spheroidal galaxies, only 10 objects remained. Their main characteristics (taken from original articles [19--22] or the NED) are listed in Table 2, where the contents of the columns do not require further explanation.

\vspace{6cm}

\begin{table}
\caption{List of presumed isolated dSph galaxies in the volume $V_{LG} < 3500$ km/s}
\begin{tabular}{lccclcl}\hline
Name &    RA   (J2000.0) Dec &     $a^{\prime}$&  $b/a$ &   mag&   $A_B$& Other name  \\ \hline
 (1)     &    (2)       &    (3) & (4) &   (5) &  (6)     &     (7)\\
\hline
KKH  9   & 02 12 54.9  +32 48 54 &  1.1&  0.91 &  16.6&  0.42 &   UGC 1703   \\
KKs     5   & 03 14 07.2  -37 59 33 &  1.2&  0.58 &  16.7&  0.07 &   ESO 300-20 \\
KKH 65   & 10 51 59.2  +28 21 45 &  0.7&  1.00 &  17.0&  0.08 &   BTS 23     \\
KKH 66   & 11 02 22.3  +70 15 53 &  1.0&  0.60 &  18.0&  0.10 &   KDG 74     \\
KKH 67   & 11 23 03.5  +21 19 37 &  1.2&  0.75 &  16.8&  0.10 &              \\
KK 180   & 13 04 29.9  +17 45 32 &  1.4&  0.50 &  16.7&  0.10 &   LSBC D575-8\\
KK 227   & 13 56 10.1  +40 18 12 &  0.7&  0.64 &  17. &  0.04 &              \\
KKR   8   & 14 19 14.8  +03 07 26 &  1.3&  0.62 &  18.5&  0.15 &              \\
KKR   9   & 14 27 05.0  +22 41 25 &  0.9&  0.67 &  19  &  0.29 &              \\
KK 258   & 22 40 43.9  -30 48 00 &  2.1&  0.62 &  17.4&  0.06 &    ESO 468-20\\
\hline
\end{tabular}
\end{table}

Notes for Table 2:

 KKH 9: According to the NED, this is an elliptical dwarf dE. On the SDSS image it appears to be dSph; no neighbors within the range of 500--2500 km/s found in a region with a radius of 120$^{\prime}$.

 KKs 5: Appears to be dSph on DSS-1 and dSph/dIr on DSS-red images. The object LSBG F~300-1 (dE) with $V_{LG}$ = 1456 km/s is at a distance of 37.4$^{\prime}$ and if KKs 5 has a radial velocity close to that of LSBG F 300-1, then its linear size should be $A\sim7$ kpc, which is not typical for spheroidal dwarfs. Thus, we do not accept the calculated mutual distance ($\sim220$ kpc) and we consider KKs 5 to be an isolated object. There is also the object FCCB 035 with $V_{LG} = 871$ km/s at a distance of $\sim140$ kpc. If they have similar radial velocities, then the linear diameter is $A\sim4$ kpc.

 KKH 65: On the SDSS survey image it appears to be a dSph galaxy with a very low surface brightness (VLSB). Its membership in the NGC 4314 group implies an atypical linear diameter ($> 4$ kpc).

 KKH 66: In the DSS-red, blue, it shows up as a  dwarf spheroidal galaxy. The nearest galaxy, GCGC 333-63 has $V_{LG} = 1438$ km/s. If the radial velocities are similar, their mutual distance is roughly 450 kpc and the linear diameter $A\sim5.7$ kpc (unrealistically large for dSph). 

KKH 67: On the SDSS survey image it shows up as a typical dSph, VLSB galaxy. The dwarf galaxy LSBC D570-7 with $V_{LG} = 713$ km/s is at an angular distance of 29.8$^{\prime}$. If they are a physical pair, then the linear size of KKH 67 is $A=3.5$ kpc and the projected distance is 85 kpc. We consider KKH 67 to be nominally isolated.

 KK 180: On the SDSS survey image it appears as a typical dSph galaxy. If we assume that KK 180 is at the same distance as KK 173 with $V_{LG}=952$ km/s,  we obtain a linear diameter of $\sim5.3$ kpc and a mutual distance of $\sim320$ kpc for KK 180. If, on the other hand, KK 180 belongs to the periphery of the NGC 4826 group, then its diameter $A\sim2.2$ kpc and its nearest neighbor, at a distance of $\sim220$ kpc, turns out to be KDG 215 with $V_h = 419$ km/s. We consider this galaxy to be nominally isolated.

 KK 227: In a field of radius 120$^{\prime}$ there are many galaxies with $V_h\sim2300$ km/s, the closest of which is NGC 5571 with $V_{LG} = 2637$ km/s. If KK 227 is a physical companion of NGC 5571 (at a projected distance of 117 kpc), then its linear diameter of $\sim7.3$ kpc is too large for a typical dwarf spheroidal galaxy.

 KKR  8: On the DSS-1 blue image it appears as a typical dSph galaxy. The nearest galaxy is NGC 5576 with $V_{LG} = 1437$ km/s (at a projected distance of 146 kpc), a member of the group NGC 5666. But if KKR  8 belongs to this group, then its linear diameter $A = 7.4$ kpc is too large for dSph.

 KKR  9: On the DSS-1 blue image it appears as dSph, VLSB. We assume that the nearest neighbor is a member of the Local volume, DDO 187. But if these galaxies are a physical pair (107 kpc), then the linear size of KKR  9 is $A = 0.6$ kpc. Furthermore, within a region of radius 85 $^{\prime}$ there are several galaxies with radial velocities of $\sim1100-1200$ km/s; the closest of all (390 kpc) is a magellanic-type dwarf galaxy CGCG 133-84 which seems not to be  gravitationally bound with KKR  9. Note that, by its morphology, object KKR  9 is similar to the known isolated dSph galaxy KKR  25, which is located at a distance of 1.9 Mpc.

KK 258: On the DSS-1 survey image this shows up as a dSph/dIr galaxy and on DSS-red, as dSph. If we assume that
the galaxy NGC 7361 with $V_{LG} = 1130$ km/s (at a projected distance of 250 kpc) is a physical neighbor, then the linear diameter of KK 258 is equal to 11 kpc, which is unrealistic. We note that the radial velocity of KK 258 given in LEDA, $V_h = 27246$ km/s, is wrong. 

Figure 1 shows the all-sky distribution of the isolated dwarf galaxies in equatorial coordinates. The solid circles indicate dwarf galaxies of later types and the open circles mark dwarf spheroidal galaxies. It appears, that the distribution of the late-type dwarfs is not entirely uniformly random. There is an indication of some excess in the number of objects in the region centered at RA $16^h.0$, Dec = $+15$. The median velocity of these 12 galaxies is +1800 km/s. The position on the sky and the characteristic velocity suggest that this group is located on the distant edge of the Local Void [28].\begin{figure}[h]

\begin{center}
\includegraphics[width=0.8\textwidth]{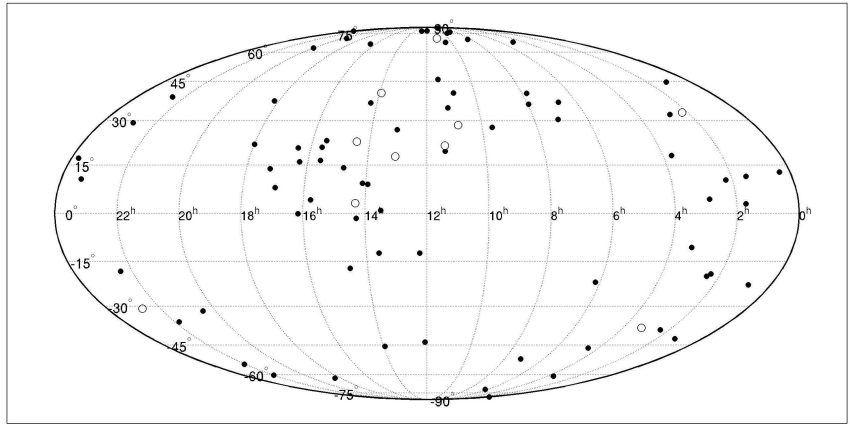}
\caption{Distribution of isolated dwarf galaxies over the sky in equatorial coordinates. The solid circles indicate late-type dwarf galaxies  and the open circles, spheroidal dwarf galaxies}
\end{center}
\end{figure}

\section{The most isolated galaxies in the Local volume}

As a supplement to the above lists (Tables 1 and 2), in Table 3 we list the galaxies in the Local volume which are the most isolated ($\Theta\leq-2.0$ ) according to the criterion of the CNG [2]. A value of $\Theta  = -2.0$ corresponds approximately to a local density of matter two orders  below the mean density. The columns in the table list the following: (1) galaxy name, (2) morphological type according to our classification, (3) corrected linear size in kpc, (4) maximum rotational velocity in km/s, (5) absolute B magnitude, corrected for Galactic and internal absorption, (6) logarithm of the hydrogen mass-to-B luminosity ratio, in solar units, and (7) the ``tidal index'' $\Theta$.
   The data in all the columns, except 2, are taken from Table 4 of the CNG. The last row of the table gives the median values of the corresponding characteristics of the galaxies.
\begin{table}
\caption{Characteristics of the most isolated galaxies in the Local volume}
\begin{tabular}{lccrrrc} \\
\hline
Name    &  Type   &  $A_B$ & $V_m$ &$M^c_B$& $\log M_{HI}/L$ & $\Theta$ \\
\hline
UGC 288   &  Im      &2.47   &19   &$-$13.82  &0.01    &$-$2.1 \\
And IV    &  Ir/Sph  &2.32   &49   &$-$12.60  &1.09   &$-$2.1 \\
IC 2038   &  Sdpec   &2.63   &50   &$-$14.42  &0.48    &$-$2.2\\
KK 49     &  Im      &1.65   &34   &$-$14.94 &$-$ 0.42  &$-$2.0\\  
ORION DW  &  Sm/Im   &6.19   &81   &$-$17.04 &$-$ 0.14   &$-$2.2\\
ESO 364-29&  Impec   &7.61   &40   &$-$16.04  &$-$0.27   &$-$2.9\\
ESO 489-56&  Impec   &0.80   &9    &$-$13.07   &$-$0.34 &$-$2.1\\ 
UGC 3476  & Im/Sm    &1.98   &38   &$-$14.27   &0.25  &$-$2.2\\
KK 55     &  Ir      &1.54   &29   &$-$13.71   &$-$0.43  &$-$2.6 \\
FG 202    & Ir       &4.63   &38   &$-$14.01   &0.38    &$-$2.0 \\
ESO 558-11& Im       &4.52   &61   &$-$16.51   &$-$0.89   &$-$2.5\\ 
UGC 3755  & Ir       &2.51   &16   &$-$14.90   &$-$0.52   &$-$2.1\\
DDO 47    & Sm/Im    &4.65   &46   &$-$15.10    &0.45   &$-$2.1\\
ESO 6-1   & Sm/Im    &2.54   &$-$  &$-$14.86      &$-$    &$-$2.4 \\
KKH 46    & Im       &1.17   &16   &$-$11.93    &0.35   &$-$2.0\\
ESO 564-30& Sm       &3.86   &68   &$-$14.27   &0.80   &$-$2.0 \\
KK 246    & Ir       &2.14   &25   &$-$12.96    &0.41   &$-$2.2\\
IC 5052   & Sd       &7.64   &83   &$-$18.23   &$-$0.49   &$-$2.2\\
\hline 
Medians   &  -       &2.51   &38   &$-$14.34    &0.01  &$-$ 2.1\\
\hline
\multicolumn{7}{l}{Notes for Table 3:}\\
 \multicolumn{7}{l}{KK 49 and Orion is located in the region of strong Galactic }\\
 \multicolumn{7}{l}{absorption ($A_B = 2.46^m$ and $3.16^m$,respectively); }\\
 \multicolumn{7}{l}{IC 5052 is not a LSB dwarf galaxy.} \\
\end{tabular}
\end{table}
The galaxies in Table 3 were checked independently as to whether they meet the isolation conditions used in this paper. It turned out that only the galaxies UGC 3476, FG 202, UGC 3755, ESO 6-1, and KK 246 met these conditions fully. UGC 288 and And IV form a wide ``pair'' with a projected mutual distance of about 460 kpc. The galaxy IC 2038 forms a close pair with IC 2039 which is a member of the group NGC 1566. The galaxies KK 49 and Orion have a difference of $<100$ km/s in their radial velocities and have a projected distance on the order of 200 kpc. The peculiar loop shaped galaxy ESO 364-29 has close (up to 150 kpc) neighbors AM 0605-341 and NGC 2188 with mutual radial velocity differences $< 50$ km/s. ESO 489-56 has a rather high surface brightness and is at a distance of about 160 kpc from ESO 490-17 (radial velocity difference 12 km/s). The galaxy KK 55 has a neighbor, NGC 2310 (radial velocity difference 321 km/s and mutual separation $\sim400$ kpc), and can be regarded as isolated, the same can also be said of the galaxy KKH 46, which has a faint neighboring galaxy at a distance of 270 kpc. The galaxy ESO 558-11 has a neighbor, HIPASS J0705-20 (radial velocity difference 34 km/s, mutual separation 130 kpc), while
DDO 47 is a member of a close pair with CGCG 37-33 (radial velocity difference 7 km/s, mutual distance 11 kpc). The galaxy ESO 564-30 can be regarded as sufficiently isolated; its nearest neighbor ESO 565-3 is located at a distance of $\sim290$ kpc (velocity difference 61 km/s), and the bright galaxy NGC 2835 (velocity difference 118 km/s) is located at a distance of $\sim450$ kpc. We are providing this detailed comparison in order to demonstrate the arbitrariness of the criterion for selecting faint dwarf galaxies as ``truly'' isolated when the mutual spatial separations and spatial velocities of the galaxies are unknown.

\section{Discussion}

The procedure of clusterizing galaxies in the volume of the Local supercluster and its surroundings ($V_{LG} < 3500$ km/s) using a stronger criterion of dynamic isolation has led to creation of the LOG catalog [11], which contains 520 galaxies at galactic latitudes $\mid b\mid > 15^{\circ}$ . Their relative number is only 4\% of the N = 10914 galaxies in this volume. Most of the LOG-galaxies are normal spiral galaxies of late morphological types. If we choose only galaxies of types Ir, Im, Sm, and BCD (T = 9, 10) from this catalog, their total number is N (9,10) = 87. This subsample of isolated galaxies included 16 of the objects in Table 1: UGC 199, ESO 243-50, UGC 2352, KDG 32, HIJ0342+1808, UGC 3672, UGC 4117, KKSG 47, UGC 9875, KKR  21, KKR  26, KKR  30, KKR  34, KKR  75, UGC 12070, UGC 12771. Thus, about 20\% of the dwarf galaxies simultaneously  pass tests for physical isolation according to the two different isolation criteria. We believe this fraction is quite high. As a comparison, we note that a recent study [29] examined the effect of the surroundings on the HI and optical properties of 101 galaxies with absolute magnitudes $M_r - 5\log h_{70} > -16$ from the catalog [30] compiled from the SDSS survey data. The average absolute magnitude of the galaxies is $<M_r> = -15.4$ and the average depth of their sample is $30 h^{-1}_{70}$ Mpc, both similar to our values. It was found [29] that 65\% of the galaxies in this sample have no neighbors in the RC3 [31] with $M_r - 5\log h_{70} < -19$ at a projected distance of $500 h_{70}$ kpc and a radial velocity difference within the range of $\pm300$ km/s. In this paper we have used a similar method for identifying isolated galaxies. Here the initial samples relied on various optical surveys. Furthermore, in [29] candidates were selected in terms of absolute magnitude, and in our work, mainly in terms of their morphological type and surface brightness. Probably, this is the case why there are no common objects in our list and in [29].

 It should be noted that the isolated irregular galaxies in Table 1 do not differ significantly in size, luminosity, and amplitude of their internal motions from other galaxies of the same type in groups, or from the gas-rich dwarf galaxies examined in  [32]. However, isolated late-type dwarf galaxies do differ significantly from non-isolated galaxies with an enhanced neutral hydrogen content per unit luminosity. Thus, in the Local volume, the population of dwarf galaxies detected in the HI line has a median value of $M(HI)/L_B = 0.8$ in solar units, while the isolated objects in Table 1 have a median of $M(HI)/L_B = 2.0$. Evidently, the absence of external tidal perturbations in the isolated dwarf galaxies does not favor the rapid expenditure of the gas contained in them on star formation processes. 

The apparent magnitudes and surface brightnesses of the presumed isolated dSph galaxies are similar to those for spheroidal galaxies located in the nearby volume. So far, we are aware only of two cases of isolated dSph: KKR25 
and APPLES1, which have distances of 1.9 and 8.5 Mpc and absolute magnitudes $M_B = -9.9$ and --8.3, respectively. The discovery of new dwarf galaxies populated only by old stars and lying beyond  of groups and clusters regions, will be of fundamental interest for modern cosmological scenarios of galaxy formation.

 In this paper we have used the DSS (http://archive.eso.org/dss) and SDSS (http://www.sdss.org) digital sky surveys, as well as the HYPERLEDA (http://leda.univlyon1.fr/) and NED (http://nedwww.\\ ipac.caltech.edu) databases. This work was partially supported by grants from RFFI, No. 09-02-90414-Ukr-f-a-i, MON Ukrainy, No. F28.2/059.

\bigskip

{\bf REFERENCES} \\

\bigskip

1. Proceedings of IAU Conference ``Galaxies in Isolation: Exploring Nature vs. Nurture'', May 2009, Granada, Spain, ASP Conf. Ser. 421 (2010). 

2. I. D. Karachentsev, V. E. Karachentseva, W. K. Huchtmeier, and D. I. Makarov, Astron. J. 127, 2031 (2004) (CNG).

 3. F. Zwicky, E. Herzog, M. Karpowich, C. T. Kowal, and P. Wild (1961-1968), Catalogue of Galaxies and Clusters of Galaxies, California Institute of Technology, Pasadena, I-VI (CGCG).

 4. P. Nilson, Uppsala General Catalogue of Galaxies, Uppsala Astron. Observ. 6, 1 (UGC) (1973). 

5. E. Lauberts, The ESO/Uppsala Survey of the ESO (B) Atlas, Munich: ESO (1982).

 6. V. E. Karachentseva, Soobshch. SAO 8, 3 (1973). 

7. I. D. Karachentsev, Astron. Astrophys. Trans. 6, 1 (1994).

 8. I. D. Karachentsev and D. I. Makarov, Astrophys. Bull. 63, 320 (2008).

 9. D. I. Makarov and I. D. Karachentsev, Astrophys. Bull. 64, 24 (2008). 

10. D. I. Makarov and I. D. Karachentsev, Mon. Notic. Roy. Astron. Soc. 412, 2498 (2011).

 11. I. D. Karachentsev, D. I. Makarov, V. E. Karachentseva, and O. Y. Melnyk, Astrophys. Bull., 66, 1 (2011) (LOG) .

 12. V. E. Karachentseva and M. E. Sharina, The Catalogue of low surface brightness dwarf galaxies, Comm. Spec. Astrophys. Obs. 57, 3 (1988). 

13. B. Binggeli, M. Tarenghi, and A. R. Sandage, Astron. Astrophys. 228, 42 (1990) (BTS). 

14. C. D. Impey, D. Sprayberry, M. J. Irwin, and C. D. Bothum, Astrophys. J. Suppl. Ser. 105, 209 (1996).

 15. J. M. Schombert, R. A. Pildis, and J. A. Eder, Astrophys. J. Suppl. Ser. 111, 233 (1997).

16. A. B. Whiting, G. K. T. Hau, and M. Irwin, Astrophys. J. Suppl. Ser. 72, 245 (2002). 

17. T-X. Thuan and P. O. Seitzer, Astrophys. J. 231, 237 (1979).

 18. S. E. Schneider, T-X. Thuan, C. Magri, and J. E. Wadiak, Astrophys. J. Suppl. Ser. 72, 245 (1990). 

19. V. E. Karachentseva and I. D. Karachentsev, Astron. Astrophys. Suppl. Ser. 127, 409 (1998) (KK).

 20. V. E. Karachentseva, I. D. Karachentsev, and G. M. Richter, Astron. Astrophys. Suppl. Ser. 135, 221 (1999) (KKR). 

21. I. D. Karachentsev, V. E. Karachentseva, and W. K. Huchtmeier, Astron. Astrophys. 366, 428 (2001) (KKH).

 22. V. E. Karachentseva and I. D. Karachentsev, Astron. Astrophys. Suppl. Ser. 146, 359 (2000) (KKs).

 23. I. D. Karachentsev, V. E. Karachentseva, A. A. Suchkov, and E. K. Grebel, Astron. Astrophys. Suppl. Ser. 145, 415 (2000) (KKSG).

 24. W. K. Huchtmeier, I. D. Karachentsev, and V. E. Karachentseva, Astron. Astrophys. 322, 375 (1997). 

25. W. K. Huchtmeier, I. D. Karachentsev, V. E. Karachentseva, and M. Ehle, Astron. Astrophys. Suppl. Ser. 141, 469 (2000). 

26. W.K.Huchtmeier, I. D. Karachentsev, and V. E. Karachentseva, Astron. Astrophys. 377, 801 (2001).

27. W.K.Huchtmeier, I. D. Karachentsev, and V. E. Karachentseva, Astron. Astrophys. 401, 483 (2003). 

28. R. B. Tully, Nearby Galaxy Catalog, Cambridge University Press (1988).

 29. M. Geha, M. R. Blanton, M. Masjedi, and A .A. West, Astrophys. J. 653, 240 (2006). 

30. M. R. Blanton, D. J. Eisenshtein, {D. V. Hogg, et al., Astrophys. J. 631, 208 (2005).

 31. G. de Vaucouleurs, A. de Vaucouleurs, H. G. Corwin, et al., Third Reference Catalogue of Bright Galaxies, I-III, SpringerVerlag, Berlin, Heidelberg, New York (1991) (RC3).

 32. R. F. Minchin, M. J. Disney, Q. A. Parker, et al., Mon. Notic. Roy. Astron. Soc. 355, 130 (2004).

\end{document}